\documentclass[fleqn,usenatbib]{mnras}

\usepackage{aas_macros}
\usepackage{cleveref}
\usepackage{graphicx}	   
\usepackage{amsmath}	   
\usepackage{amssymb}	   
\usepackage{multicol}      
\usepackage{color}
\usepackage[utf8]{inputenc}
\usepackage[T1]{fontenc}
\usepackage[english]{babel}
\usepackage{bm}		       
\usepackage{listings}
\usepackage[skins,theorems]{tcolorbox}
\usepackage{array}
\usepackage{booktabs}
\usepackage{ulem}
\usepackage{lipsum}
\usepackage{multirow} 
\usepackage{placeins}

\newcommand{\VEC}[1] {{\boldsymbol{{ #1}}}}

\definecolor{amber}{rgb}{1.0, 0.49, 0.0}

\title{AGN dichotomy beyond radio loudness: a Gaussian Mixture Model analysis}

\author[Pedro~P.~B.~Beaklini]
{Pedro~P.~B.~Beaklini,$^{1,2}$
Allan~V.~C.~Quadros,$^{3,4}$
Marcio~G.~B.~de~Avellar,$^{5,6}$
\newauthor
Maria~L.~L.~Dantas,$^{1}$
Andr\'e~L.~F.~Can\c{c}ado$^3$\\
$^1$Instituto de Astronomia,
Geof\'isica e Ci\^encias Atmosf\'ericas, Universidade de S\~ao Paulo\\
Rua do Mat\~ao 1226, 05508-090, S\~ao Paulo/SP, Brazil.\\
$^2$National Radio Astronomy Observatory\\ 1003 Lopezville Road, Socorro, NM 87801, United States of America \\
$^3$Departamento de Estat\'istica, Universidade de Bras\'ilia, UnB\\
Campus Universit\'ario Darcy Ribeiro, Sgan, Asa Norte, 70910-900, Bras\'ilia-DF, Brazil \\
$^4$Department of Statistics, Kansas State University\\
101 Dickens Hall, 1116 Mid-Campus Drive N., Manhattan KS 66506-0802, United States of America\\
$^5$Instituto Tecnol\'ogico de Aeron\'autica, CTA \\ Pra\c ca Marechal Eduardo Gomes, 50, Vila das Ac\'acias, 12228-900, S\~ao Jos\'e dos Campos/SP, Brazil \\
$^{6}$ Departamento de F\'isica, Universidade Federal de S\~ao Paulo\\ Rua S\~ao Nicolau 210, 09913-030, Diadema/SP, Brazil 
}

\begin{document}

\maketitle

\begin{abstract}
Since the discovery of Quasi-stellar Objects (QSOs), also known as quasars, they have been traditionally subdivided as radio-loud and radio-quiet sources. Whether such division is a misleading effect from a highly heterogeneous single population of objects, or real has yet to be answered. Such dichotomy has been evidenced by observations of the flux ratio between the optical and radio emissions (usually $B$-band and 5 GHz). Evidence of two populations in quasars and samples of a wide diversity of AGNs has been accumulated over the years. Other quantities beyond radio loudness also seem to show the signature of the existence of two different populations of AGN. To verify the existence of a dichotomy through different parameters, we employed a soft clustering scheme, based on the Gaussian Mixture Model (GMM), to classify these objects simultaneously using the following parameters: black hole mass, colour and $R$ loudness index, as well as the usual radio and $B$-band luminosity. To investigate whether different kinds of AGNs manifest any population dichotomy, we applied  GMM to four independent catalogues composed of both optical and radio information. Our results indicate the persistence of a dichotomy in all datasets, although the discriminating power differs for different choices of parameters. Although the Radio Loudness parameter alone does not seem to be enough to display the dichotomy, the evidence of two populations of AGNs could persist even if we consider other parameters. Our research suggests that the dichotomy is not a misleading effect but real.

\end{abstract} 


\section{Introduction}
\label{int}

It is well known that there are different types of AGNs and that most of the diversity can be explained as a consequence of the angle between the source structures and the line of sight, following the AGN unification schema \citep[e.g.,][]{hol92,ant93,urr95,pet97,ho08}. However, it is also well known that there are some intrinsic differences between the AGN like the core or lobe dominated Fanaroff–Riley galaxies \citep{fan74}, the black hole spin \citep{mod98,bre13,bar19,una20}, the accretion rate of the central engine \citep{bia02,kor08,dal18}, and the radio to optical luminosity relation \citep{kel66,mur77,con80,kel89}. The last point is known in the literature as the radio-loud and radio-quiet quasar dichotomy, being the first evidence of the existence of two inherently different populations of AGNs \citep[see][]{pet97}. 

The existence of a dichotomy involving the radio emission of quasars is a long-lasting question since the discovery of the Quasi-Stellar Radio Sources, known as quasars, in the '60s. From the total well-known Quasi-Stellar Objects (QSOs) sample at that time,  almost 10\% of them were detected at radio wavelengths and classified as quasars \citep{chi64,kel66,sha66,kat73,fan77}, and only a few of them were classified as radio-loud sources. These peculiarities indicate that radio properties could be fundamental to classify active galactic nuclei (AGNs), which led to the first division between Radio-Loud and Radio-Quiet objects \citep[see][]{kel89}. However, whether this disjunction is physical or statistical is unclear \citep[see, e.g.,][and references therein]{kel16}.

\citet{kel89} defined the radio loudness parameter ($R$) as an indicator of whether the source is radio-quiet or loud. The authors suggested a comparison between the radio and optical $B$-band fluxes; a source was defined as radio-loud (hereafter RL) when the radio flux was higher than the $B$-band, and radio-quiet (hereafter RQ) otherwise. It is worthwhile noticing two aspects of analysing BL Lac and flat-spectrum radio quasars (FSRQ): (i) the beaming effect increases in different ways for the fluxes in radio and optical wavelengths; (ii) opacity effects are a source of biases \citep[for instance, see][]{bla79,urr95,fal96a}. By taking these issues into account, the $R$ parameter is the commonly adopted criterion used to classify a source as radio-loud or quiet. Although it is a very useful criterion, other works suggest different criteria to look for a more evident dichotomy \citep[e.g.,][]{van15}.

It is essential to discuss the meaning of a real dichotomy. Notably, some AGNs are bright at radio wavelengths while others are not \citep[e.g.,][]{chi64,sce79,smi80,sra80,str80}. Naturally, this statement is also valid at other wavelengths. It is also well known the existence of external factors that increase the continuum emission that depends on the jet position relative to the line of sight; the well-known beaming effect is one of them \citep{ree66,bla79}. Therefore, the question to be addressed is: is there a continuous change of the brightness throughout the AGN population? Is this evident only when we look at the number of quasars versus the radio loudness parameter? Or are there other types of AGNs where a dichotomy could be seen in other parameter spaces? We can also pose these questions in another way. Can other parameters evidence a dichotomy?

First, let us take into account only the radio loudness dichotomy (that we will call hereafter of the traditional RL/RQ dichotomy). In general, two main approaches try to find the existence of two populations. The first one is to search directly for a gap in the bright luminosity and the $R$ parameter space. The second is to try to determine the distribution of radio flux between RQ and RL sources to show that they are not compatible with just one general distribution \citep[e.g.,][]{str80,fal96b,whi00,cir03a,sik07,bro11,mah12,kel16}. Both are limited by a two-parameter analysis, although \citet{sik07} and \citet{bro11} also used the central source mass to investigate the bimodality.      

In the context of the first approach, \citet{sik07} found two parallel tracks in the luminosity plot. We highlight that the authors constructed a catalogue with sources of different types, expanding the original idea of quasar dichotomy: Broad-line radio galaxies, radio-loud quasars, Seyferts, LINERs, quasars, and FR I galaxies. Using the same catalogue, but considering only the core flux, \citet{bro11} used the black hole fundamental plane \citep{mer03,fal04} to investigate how the mass could affect the traditional RL/RQ dichotomy. The authors concluded that the gap previously found by \citet{sik07} was still present after the mass correction, but not as evident as before. 

On the other hand, observations made by the Faint Images of the Radio Sky at Twenty survey \citep[FIRST, ][]{bec95,whi97,whi00} provided an extensive database. The instrument used was the Very Large Array (VLA) at 1365 and 1435 MHz. It provided flux measurements of a large number of sources that would be ideal for verifying the existence of different populations. However, the data from the FIRST catalogue did not reveal two different populations of radio sources \citep{whi00,lac01,cir03a,cir03b,wal05,raf09,sin11,bon13}. Besides, analyses based on ATCA (Australia Telescope Compact Array) and the ROSAT (Röntgensatellit) All-Sky Survey did not find evidence of two intrinsically different classes of QSOs \citep{mah12}.

\citet{kel16} pointed out that the lack of detection on the FIRST catalogue could be a consequence of the large antenna beam at its frequency. In other words, the antenna detected not only the core emission but also the lobe emission, which could introduce noise to the distribution. Nevertheless, the authors also argued that the X-ray QSO catalogue \citep{mah10} favoured blazars and that the ATCA sensibility could hide the RQ population. 

\citet{kel16} have also used VLA observations of SDSS. They used the QSOs sample presented by \citet{kin11} at 6 GHz, which was based on the compilation by \citet{sch10}. The authors restricted their data in a very narrow redshift interval between z = 0.2 and z = 0.3, which made it impossible to confirm whether the redshift evolution exists. However, they revealed a bimodal distribution not found in \citet{kin11}. We highlight that the authors confirmed the definition of the traditional dichotomy, which is well-described through the radio loudness parameter and in a sample formed only by quasars.

Now, let us go through other parameters. In an attempt to perform a multi-dimensional analysis and try to solve if a dichotomy could be seen in other wavelengths, \citet{sul15} used the {\it eigenvector correlation} space considering line parameters in optical and UV range beyond the X-ray photon index \citep{sul00}. The authors found that radio-loud sources are restricted to a small group compared to the whole population of radio quasars. A potential cause for this cleavage was suggested to be quasar evolution \citep{sch00,bes12}. Some authors considered quasar luminosity evolution when searching for dichotomy \citep{sin13}. \citet{ret17} investigated the existence of quasars' spatial clustering and detected significant correlations with redshift for RL quasars at $z \leq 2.3$, which could provide valuable information about quasar evolution. The dichotomy in quasars seems to be a general property, and not only a consequence of the radio emission. 

In the context of a multi-parametric analysis, \citet{sin13} described the luminosity evolution of radio and optical emission of quasars combining Sloan Digital Sky Survey (SDSS) sources and the FIRST catalogues. The authors applied a multi-dimensional correlation method of \citet{efr92} to take into account flux limitations and sample restrictions. They did not find evidence for bimodality, but this could be a consequence of the FIRST limitations mentioned above.

To evaluate the extent of a dichotomy among AGNs, we studied a set of features from different papers and catalogues by using a probabilistic methodology. Therefore, we employed a Gaussian Mixture Model \citep[see][]{des17} to this multi-parametric space in order to: (i) test whether a dichotomy exists beyond the radio flux; (ii) if so, to find the best combination of parameters to discriminate them, and (iii) search if the dichotomy is presented even in samples with more AGN diversity than quasars. In this sense, we are not restricting to the radio emission dimension but taking an overall look at the possible dichotomies in AGNs.
 
Thus, in summary, we are searching for a dichotomy along many possible surfaces in the space of quasar parameters, only one of which is the traditional RL/RQ dichotomy, which has received a lot of previous attention in the literature.

In the next section, we present the samples we used. In section \ref{gmm}, we discuss the importance of this multi-parameter analysis in astronomy and the method we used, while in section \ref{res}, we show our results. We discuss our results in section \ref{dis}, and we conclude in section \ref{conclusion}.


\section{Data Retrieval}
\label{cat}

The multi-dimensional analysis of the radio sources' emissions can help to determine the reliability of an AGN dichotomy. Even the traditional RL/RQ dichotomy is still under debate. Different samples and analyses point many times to contradictory results. In this work, we will not compile a new database since there are already many of them in the literature, as pinpointed in section \ref{int}. Instead, we will use different samples to investigate how the dichotomy appears in other observational parameters.

In other words, we did not compile any new database; we used in this work the databases compiled and treated by the authors cited in Table 1. We did not modify or treat any data; we only normalised the data in the way described below.

We now have different questions to address since we are working with parameters other than the radio and the B band luminosities and with a broad class of AGNs. Is the dichotomy a general property? Do the two populations appear when we use a comprehensive set of parameters? Does the dichotomy appear through different classes of AGN beyond quasars?

The main problem of thinking in a multi-dimensional or multi-parameter analysis is to define which parameter space is the best one to investigate. In an ideal scenario, we would have a big catalogue with several well-determined parameters, but this is not the case. The catalogues containing SDSS information, for example, do not have the mass of the central source for most of the sources, while some objects of the FIRST catalogue do not have an optical counterpart. Then, choosing a sample is choosing the parameters worthy of investigating.

Our goal is to introduce the multi-dimensional probabilistic analysis in the discussion. We selected four datasets used in previous works as follows.

\begin{enumerate}
    \item Original database compiled by \citet{sik07};
    \item The modified version of the \citet{sik07} database by \citet{bro11};
    \item The sample of \citet{sch10} and \citet{kin11} also analysed by \citet{kel16}; 
    \item The FIRST data used by \citet{whi00}. 
\end{enumerate}

In the following, we briefly describe each sample individually.

\paragraph*{Dataset D1 (5 GHz, $B$-Band, Mass, $R$):} There are 198 objects in this sample. The authors made use of the following parameters: 5 GHz radio flux, $B$-band, mass, and Radio Loudness parameter. The disadvantages of including the black hole mass, as performed by \citet{sik07}, is the low number of well-known sources; it limits the total number of entries in the catalogue. Nonetheless, the authors correctly excluded blazars because of the boosting of their emission. Five classes of sources were obtained from different catalogues considering radio emission \citep{smi83,era94,era03}, radio structure \citep{woo02,cao04,kha04}, and Seyferts and LINERs \citep{ho01,ho02}. \citet{sik07} discussed the details of each sub-sample. This sample provides a space of parameters containing Radio and optical luminosity, the $R$ parameter, and the black hole mass, with a total of 198 sources. 

\paragraph*{Dataset D2 (5 GHz core, $B$-Band, Mass, $R$):} There are 198 objects in this sample. The authors made use of the following parameters: 5 GHz radio core flux, $B$-band, mass, and radio core Loudness parameter. \citet{bro11} also used this compilation to analyze the influence of the black hole's fundamental plane on the dichotomy. However, \citet{bro11} applied a correction on the original sample radio flux to guarantee that the used flux corresponded to the core emission, which changes not only the radio luminosity but also the $R$ parameter.

\paragraph*{Dataset D3: (Offset, 6 GHz, $B$-Band, $R$)} There are 178 objects in this sample. The authors obtained the following parameters: 6 GHz Radio Flux, B-Band, Radio Loudness parameter, and displacement between radio and optical emission. From the sample analysed by \citet{kel16} and compiled by \citet{kin11}, the authors used only a small range of redshift (between 0.2 and 0.3) to avoid the influence of evolutionary effects and used a rigorous selection criterion based on colours, magnitude (which excludes low luminosity AGNs), and the presence of broad lines (at least one). The final list had 178 SDSS quasars. The sample provides information about the luminosities at radio and I bands, the offset between radio and optical emission, the $R$ parameter, and the redshift. 

\paragraph*{Dataset D4: (1.4 GHz,$B$-Band, $R$-Band, colour ($B$-$R_{b}$))} There are 636 objects in this sample. The authors obtained the following parameters: $1.4$ GHz Radio Emission, $B$-band, $R$-band (representing as $R_{b}$), and$R$Loudness parameter. The fourth and last sample we used in this work was based on the FIRST catalogue. No dichotomy has been reported in this dataset, although reasons for the lack of detection exist, as discussed in section \ref{int}. Our goal is to verify whether a probabilistic multivariate method like GMM can indicate some new information even at this beam resolution.

\vspace{0.5cm}

We stress that all samples have already been accounted for the K-correction and that sources with strong beaming effects were not considered. Nonetheless, \citet{fal96a} showed that the boosting effect could affect even RQ sources, but \citet{bro11} argued that it is not a crucial matter for the \citet{sik07} sampling, while \citet{kel16} ruled out the Doppler boosting as the origin of the dichotomy.

For all samples, we first took the logarithm of the values of each parameter. Then, we scaled the new values from 0 to 1 by the formula below, where $min$ and $max$ represent the minimum and maximum value of each parameter:

$$
f(x) = \frac{(x-min)}{(max-min)}.
$$

We emphasise that the scaling does not affect the main results. This procedure's objective is to avoid that any local minimum in the statistical method blurs the result. We also used this normalised scale on our plots.

To investigate the existence of a dichotomy, we also need to discuss a suitable approach concerning the outliers of each sample. Notice that since we are trying to identify a supposed second population, this second population could be precisely a group of outliers. After removing entries below the 5th percentile and above the  95th percentile, the number of sources may decrease significantly; thus, the outliers are the points that fall outside the 5-95 percentile range for any parameter. We present and discuss both results -- with and without the outliers -- and deliberate on both results.

At this point, we must emphasize that the definitions of radio-loud and radio-quiet may not be enough to interpret the groups we find. Since we are now dealing with a multi-parameter space, the groups do not necessarily correspond to the RL or RQ groups.
 
\begin{table*}
    \centering
    \caption{Summary of samples used in this work}
    \begin{tabular}{c|l|l|c|c}
         Dataset & References & Parameters & Objects & no Outliers\\
         \hline
         \hline
         D1 (Sec. \ref{res1}) & \citet{sik07} & 5 GHz, $B$-Band, Mass, $R$ & 197 & 131\\
         D2 (Sec. \ref{res2}) & \citet{sik07,bro11} & 5 GHz (core), $B$-Band, Mass, $R_{core}$ & 197 & 131\\
         D3 (Sec. \ref{res3})& \citet{sch10, kin11} & Offset, 6 GHz, $B$-Band, $R$ & 175 & 116\\
         & \citet{kel16} & &  & \\
         D4 (Sec. \ref{res4})& \citet{whi00} & 1.4 GHz,$B$-Band, $R$-Band, ($B$-$R_{b}$) & 636 & 419 \\ 
    \end{tabular}
    \label{par}
\end{table*}{}


\section{Methodology}
\label{gmm}
This section briefly describes the methodology chosen to address the issue involving different clusters of QSOs. We employed a probabilistic approach known as the Gaussian Mixture Model (GMM) to evaluate the presence of different clusters or groups in each dataset.


A Mixture Model is defined as a probability density function comprised of a weighted summation of component densities. Therefore, in the GMM, we have a summation of Gaussian Components (GC),  in which each of these GCs indicates one potential cluster in the data \citep{McLachlan00,hastie01,mengersen2011,Murphy2012,des17, Ucci2018}. 

For a total of $k$ components in a $d$-dimensional parameter space, one can write the GMM as:
\begin{equation}
  f(x)  = \sum_{i =1}^k \zeta_i\phi(x;\VEC{\mu}_i,\VEC{\Sigma}_i),
  \label{eq:px}
\end{equation}
with mixture weights denoted by $\zeta_i$, and $\sum_{i =1}^k \zeta_i = 1$. Each of the $k$ components is described as a $d$-variate Gaussian density, fully characterised by its mean $\VEC{\mu}_i$ and covariance matrix $\VEC{\Sigma}_i$, given by:
\begin{equation}
  \phi(x;\VEC{\mu}_i,\VEC{\Sigma}_i) = \frac{1}{\sqrt{(2\pi)^d|\VEC{\Sigma}_i|}}e^{-\frac{1}{2}(x-\VEC{\mu}_i)'\VEC{\Sigma}_i^{-1}(x-\VEC{\mu}_i)}.
\end{equation}

In this scenario, to each point in the dataset, we assign a probability of membership to each GC. Thus we assign the point to the cluster to which it has the highest probability of membership after the convergence of the fitting algorithm. Fitting a GMM means estimating the maximum likelihood function of the Gaussian Mixture. The estimation is done using the Expectation-Maximization (EM) algorithm. This method consists of a data augmentation procedure with underlying data, represented by the model memberships, during the fitting procedure of a GMM \citep[]{hastie01, Dempster77,Hoboken08}.

The EM algorithm from the \textsc{r} \citep{rlanguage} package \textsc{mclust} \citep{mclust} is used to fit the Gaussian Mixture models to each dataset. Since our goal is to verify whether there is a bimodality in each type of source, we fitted solutions for $k=1$ and $k=2$ clusters.


In order to evaluate whether the results have indicated or not the presence of a dichotomy, we have employed a model selection framework along with a hypothesis testing methodology to evaluate how significant is the hypothesis of having two or more heterogeneous populations in each dataset.

First, we used the Integrated Completed Likelihood (ICL) criterion \citep[]{biernacki00} to measure each solution's goodness of fit. We were able to indicate the presence or absence of a dichotomy in each parameter space by comparing the ICL of GMMs with $k=1$ and $k=2$ groups. We used the \textit{icl()} function from the \textsc{mclust} package, where a greater ICL indicates a better model or solution.

The ICL is an alternative to the standard and widely used Bayesian Information Criterion (BIC) and better suited to our purposes in this paper. In the context of density estimation, BIC has a satisfactory performance. However, it tends to overestimate the number of clusters when employed in the model-based clustering analysis context \citep{biernacki00}. In our case, this could result in some bias towards the $k=2$ solutions. Although some authors have used it to assess the number of clusters in data \citep{biernacki00}, when some of the regularity conditions for the Normal distribution are not present, more Gaussian components are fitted to encompass the whole data or a unique cluster, thus overestimating $k$. Since our datasets do not follow a normal distribution, this reinforces our option for the ICL.

A disadvantage of the ICL is that there is no mention to the significance of the difference between ICL values when comparing two models, as \citet[]{raftery95} has provided for the BIC using the Bayes Factor framework. A difference in the BIC values from two models represents an odds ratio given by the probability that the model $B$ is correct given the data, over the probability that model $A$ is correct given the data. For example, a difference of $|10|$ between two BICs represents a $150:1$ odds ratio that the model with the smaller BIC is the better fitted one. Considering that the ICL formula contains a BIC-like approximation, we can use this number as a loose reference when comparing ICL values for two solutions.

 As a complement to the ICL criterion, we employed a hypothesis-testing framework. \citet[]{fuentes09} developed a Bayesian procedure to test for the existence of clusters in a multivariate dataset. The main goal is to test $H_0: k = 1$ vs $H_1: k > 1$ . Here, $H_0$ is the null hypothesis and denotes the existence of a unique group (no dichotomy) in the sample. The alternative hypothesis, $H_1$, denotes the possibility that there is more than one group in the data and can be simplified to $k = k_{n}$, where $k_{n}$ is the number of groups that we want to test against $k=1$. Therefore, the method allows us to test whether the data represent a single homogeneous population or two (or more) populations. Since we are interested in evaluating the existence of a dichotomy, i.e. $k_{n}=2$, we tested in favour of $H_1: k = 2$\footnote{We used the  \textsc{bayesclust} package \citep[]{gopal12}, an \textsc{r} \citep{rlanguage} implementation of the aforementioned method.}. Thus, we obtain a measure of the significance of the alternative hypothesis $k=2$. That means we can evaluate if we have sufficient evidence to reject the hypothesis of a single and homogeneous population, indicating instead the existence of two separate groups (dichotomy). 

 In the hypothesis testing framework, the significance level, $\alpha$, is also known as the probability of making a Type I error, i.e., the probability of wrongly rejecting the null hypothesis $H_0$ \citep[]{johnson06}. In our case, it means that we would reject the hypothesis of no group division ($k=1$) when, in fact, there are no discrete populations of AGNs. The $\alpha$ significance level is set \textit{a priori} and will serve as the base of comparison for our p-value, which is the probability of observing a value at least as extreme as we have observed for the test statistic under the null hypothesis distribution (i.e., assuming that $H_0$ is correct). 

We opt for a strict $1$ per cent $\alpha$ (i.e., we set $\alpha = 0.01$) level to evaluate the significance of the hypothesis testing results. We reject the null hypothesis if the computed p-value is less than $0.01$. Therefore, by setting a small value for $\alpha$, such as one percent, we are very rigorous since we would only reject the hypothesis of a single AGN population ($k=1$) if the p-value is smaller than $0.01$\footnote{The probability of observing the computed test statistic under the null hypothesis distribution ($H_0$) has to be considerably small to assume that we are actually dealing with values under the probability density of the alternative hypothesis ($H_1$).}. This is equivalent to say that the data have to provide robust evidence so that we decide to reject the hypothesis that there is only one homogeneous population of AGNs.

Therefore, our methodology consists of fitting Gaussian mixture models with $k=2$ solutions to each dataset (with and without the outliers); fitting GMMs with $k=1$ solution and comparing ICL values for $k=2$ and $k=1$ solutions; and, in order to complement the analysis, we evaluate the significance of the results by testing against $k=1$ (in favour of $k = 2$), using a hypothesis testing framework.

 
\section{Results}
\label{res}

This section describes the results of the application of the GMM to the four catalogues. Each catalogue reflects a different space of parameters, hence enabling to test the robustness of a dichotomy prevalence against different galaxy properties. 

In what follows, we describe each GMM fit. It is worth noting that we performed all analyses in the multivariate space composed by all parameters of each catalogue, but for visualisation purposes, we display the projected pair-wise two-dimensional solutions.

 \begin{figure}
  \includegraphics[width=0.865\linewidth]{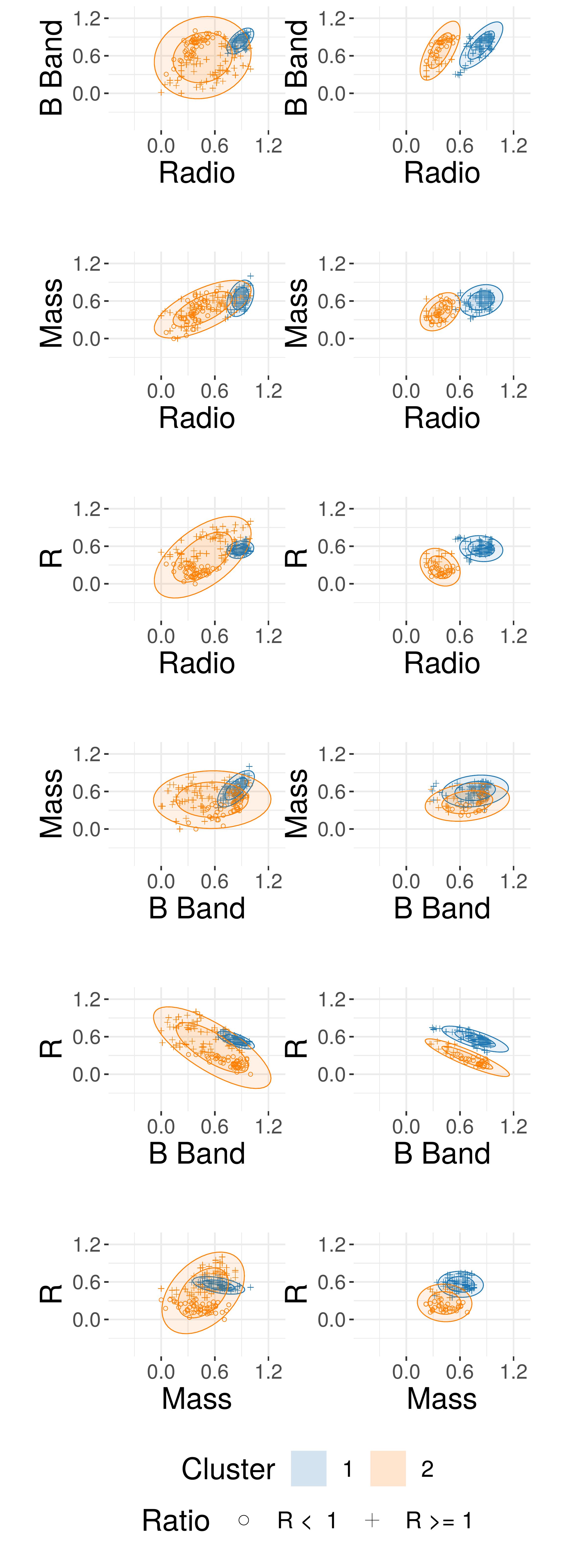}
  \caption{GMM results for the space of parameters for dataset D1 (parameters 5 GHz flux density, $B$-Band magnitude, Mass, $R$ index ), where the circles indicate a confidence interval of 95 and 68 per cent. On left: the whole sample. On right: After removing the outliers.}
  \label{ds1}
\end{figure}

 \begin{figure}
  \includegraphics[width=0.8\linewidth]{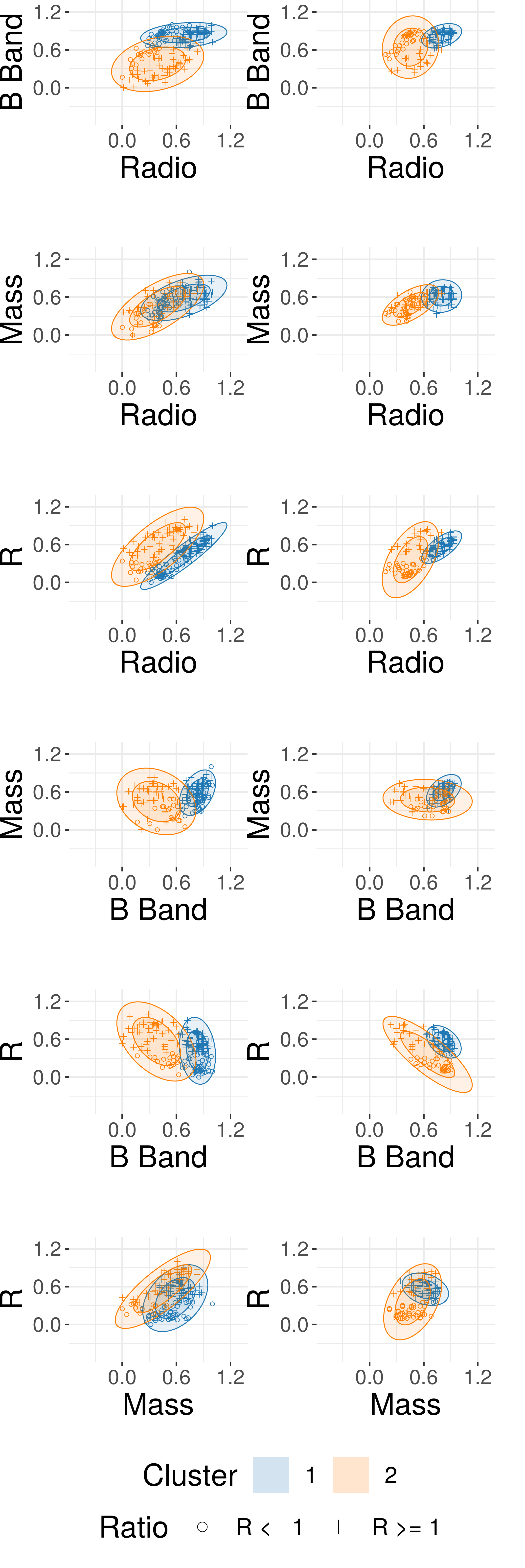}
  \caption{GMM results for the space of parameters for dataset D2 (parameters 5 GHz core flux density, $B$-Band magnitude, Mass, $R$ index ), where the circles indicate a confidence interval of 95 and 68 per cent. On left: the whole sample. On right: After removing the outliers.}
  \label{ds2}
\end{figure} 


\subsection{D1: parameter space: Radio and B Bands, Mass and R}
\label{res1}

We shall note that, because this sample presents a broader diversity of sources (FR I galaxies, PG quasars, Broad Line Radio galaxies, Seyferts, and LINERS), \citet{sik07} weighted the $R$ parameter by the Eddington Luminosity, which differs from the classical definition commonly applied to quasars.

We present the GMM fit in the left column of Figure \ref{ds1}. We use different symbols depending on the radio loudness parameter value to determine the extent to which the groups correspond with the traditional RL/RQ dichotomy: crosses indicate sources with $R>1$, and circles represent sources with $R<1$. Colours in the figures represent the best group division we found. The result is that the ICL slightly favours the presence of two groups instead of one. 

In this same Figure, one can see the 95 and 68 per cent confidence regions limited, respectively, by the outer and inner ellipses. These regions are related to the estimated population parameters after fitting the Gaussian mixtures and should not be confused with a quantile region for the sample. To illustrate their meanings, let us take the 95 per cent region as an example. It indicates that if we were to replicate the sampling from the underlying fitted distribution as many times as possible, and for each replication we compute the confidence region, then 95 per cent of the ellipses constructed this way would contain the real mean of the population. Likewise, for the inner ellipses, 68 per cent of the intervals ellipses so constructed would contain the underlying mean.

Based on the $R>1$ distribution in the groups, it would be possible to find a correspondence between the multi-parameter dichotomy and the traditional RL/RQ dichotomy:  In one group, there are 80 sources $R>1$, and no sources with $R<1$, and then this could be the correspondence with the RL population. However, the total number of $R>1$ is 143 sources. Our multi-parameter analysis indicates that 63 sources known as RL would be better classified together with the RQ, in opposition to the criterion of using the weighted $R$ parameter.

We should note that \citet{sik07} defined the $R$ parameter estimating the bolometric luminosity. We agree that this redefinition of the $R$ parameter has the advantage of expanding it to other AGNs, but maybe to use it, we need to think about the dichotomy concept more broadly. Moreover, the group classification we found seems to have a clear division at normalised radio flux of $0.75$, indicating that in our RL definition, the radio luminosity could be used as a fundamental quantity to the dichotomy without the need to compare with any other luminosity. 
Nonetheless, using the whole sample, we did not find the same dichotomy as found by \citet{sik07}, in which the sources followed two different and parallel tracks on the plane between radio and B luminosities (see left column). However, when we remove the outliers, the parallel tracks appear as a natural division between the groups. In this approach, the total sample has 131 sources. The group in blue (Figure \ref{ds1}), which has more Radio Loud sources, has a total of 46 sources. It seems very similar to the traditional RL/RQ dichotomy that \citet{sik07} have found. 

This indicates that, at least for this weighted $R$ parameter, the loudness parameter only makes the dichotomy clear if we remove the most extreme objects. In this case, the existence of two parallel tracks in all the planes is remarkable, as shown in the right column of Figure \ref{ds1}. In section \ref{dis}, we discuss the contribution of each parameter. Nonetheless, the detection of a dichotomy using GMM is compatible with the results of two AGN populations. 

In Table 2, we present the quantification of the comparison between the ICL values for $k = 2$ and $k = 1$. For dataset D1, the comparison indicates that the dichotomy is likely to be present. The p-value of the significance test against $H_0: k = 1$ corroborates the dichotomy. When we removed the outliers, the dichotomy persisted: the ICL for $k=2$ clusters is higher than the ICL computed for the GMM with $k=1$. The p-value in the table indicates there is strong evidence to reject the null hypothesis of a homogeneous population in the dataset D1  since it is smaller than the one adopted at a significance level.

 We advocate that the presence of the outliers, the most extreme objects, should be taken into account to define the division line between the two populations, but this needs further discussion.
 
\begin{table}
	\centering
	\label{tabmodsel}
	\small{\caption{Model selection \& hypothesis testing results for dataset D2. The model with greater value of ICL is favoured. The p-value is used to complement the analysis. If it is smaller than $\alpha$, i.e. $< 0.01$, there is sufficient statistical evidence to reject the hypothesis of a single AGN population.}}
	\begin{tabular}{@{}rrrcrr@{}}
		\phantom{abc} & \phantom{abc} & \multicolumn{2}{c}{ICL} & \phantom{abc} \\
		\cmidrule{3-4}
		Dataset & Outliers & $k=2$ & $k=1$ & p-value \\ \midrule
		
		D1 & included & 1846.367 & 1710.1796 & 0.004125 \\
		
		D1 & removed & 1425.617 & 1315.602 & 0.006875 \\
		
		D2 & included & 1896.847 & 1812.5682 & 0.006000 \\
		
		D2 & removed & 1408.619 & 1383.010 & 0.005375 \\
		
		D3 & included & 2237.868 & 675.7179 & 0.000125 \\
		
		D3 & removed & 2205.146 & 1795.334 & 0.000125 \\
		
		D4 & included & 17892.122 & 10090.5548 & 0.000125 \\
		
		D4 & removed & 7622.312 & 7634.255 & 0.007500 \\
		
	\end{tabular}
\end{table}

%
%
%

\subsection{D2, parameter space: Core, B Band, Mass and $\mathrm{R_{core}}$}
\label{res2}

This dataset corresponds to the dataset of \citet{bro11}, adapted from \citet{sik07}. This set of parameters is similar to the one in Section \ref{res1}, although considering the core flux instead of the total radio flux. Thus, we need to evaluate the influence of occasional contamination of the extended emission of the central source flux in the results. 
 
The results led to one group having two times more objects than the other (one with 130 and other with 67 points). It is not clear whether one of these groups could be associated with RL or RQ. Taking into account the $R$ parameter, the group with 67 members, indicated in orange on the left column of Figure \ref{ds2}, can be identified as the Radio Loud. This proportion is similar to what we have found in the previous data sample. On the plot $R$ index versus radio, we can identify the parallel tracks, although the gap is almost nonexistent. 

However, the sources with the highest radio fluxes are together with the sources with higher B values. The division seems to be remarkably evident around $0.75$ of the B normalised flux. Both, $R < 1$ and $R > 1 $, sources are present, which is expected since the $R$ index compares the radio flux with the $B$-band flux, even when the Eddington Luminosity weights $R$.

\begin{figure}
  \includegraphics[width=0.81\linewidth]{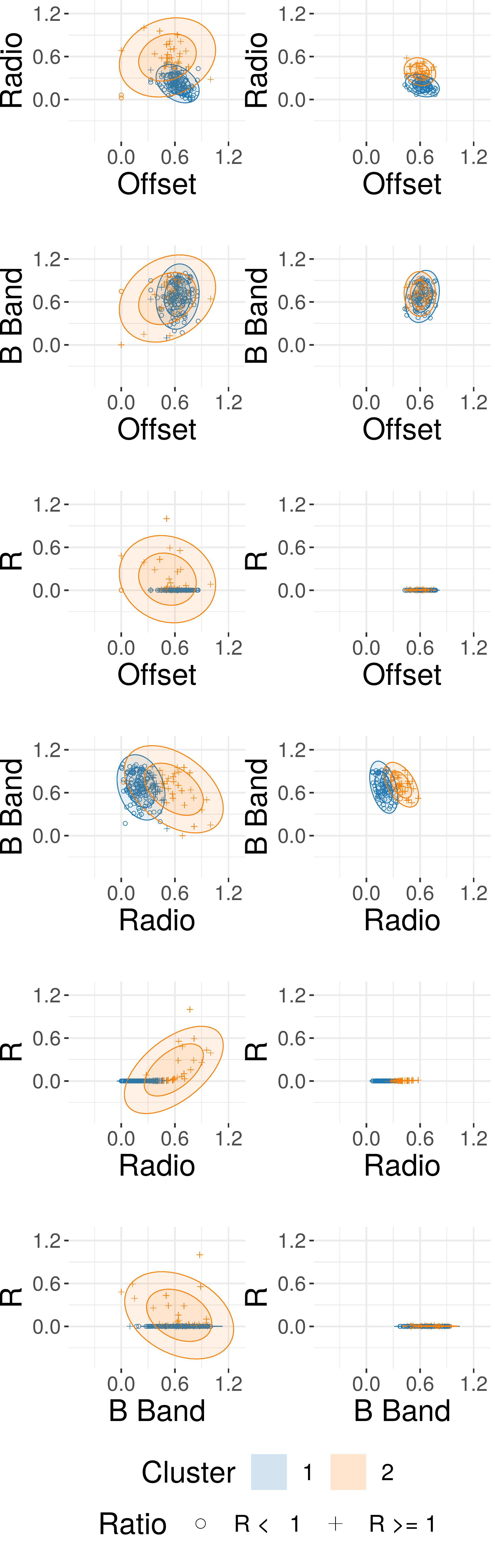}
  \caption{GMM results for the space of parameters for dataset D3 (parameters Offset between radio and optical emission, 6 GHz flux density, $B$-Band magnitude, $R$ index), where the circles indicate a confidence interval of 95 and 68 per cent. On left: the whole sample. On right: After removing the outliers.}
  \label{ds3}
\end{figure} 


\begin{figure}
 \includegraphics[width=0.85\linewidth]{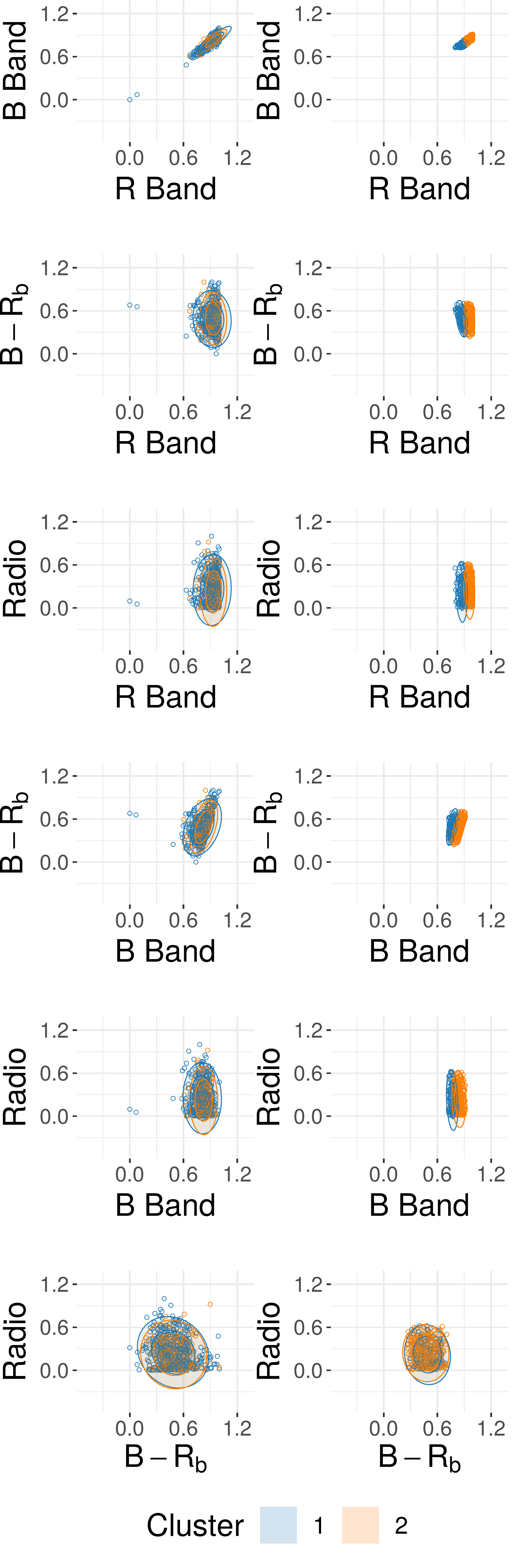}
  \caption{GMM results for the space of parameters for dataset D4 (parameters 1.4 GHz flux density, $B$-Band magnitude, $R$-Band magnitude, and ($B$-$R_{b}$) colour). The sub-index b in the colour was introduce to distinguish the radio loudness parameter from the $R$-Band magnitude. The circles indicate a confidence interval of 95 and 68 per cent. On left: the whole sample. On right: after removing the outliers.}
  \label{ds4}
\end{figure}

The proportion between the groups becomes more balanced if we remove the outliers out of the sample, but the total sample decreases to 120 sources. However, the dichotomy is less evident in this case. As discussed in Section \ref{res1}, by removing the outliers, some high luminosity radio sources are also removed. Even though we find RLs in both groups, the smaller number of sources also points to the existence of two clusters in the radio \textit{versus} optical emission chart. While not removing the outliers points to two groups with similar dispersion, now the dispersion is smaller in one of the groups. The latter group has more Radio Loud sources, which can be seen by the blue circles in Figure \ref{ds2}. In other words, even if one of the groups has more RL sources, in this sample, it is not easy to identify a correspondence between the division we found and the traditional RL/RQ dichotomy.

\citet{bro11} noted that replacing the total flux by the core emission decreases the gap between parallels tracks on the radio versus $R_{band}$ luminosity plot. In this approach, after removing the outliers, no evidence of the parallel tracks is found. 

One could ask if we can also use the lobe emission, defined as the difference between the total and the core emission since we have both. In principle, the lobe flux should not be a vital feature to the dichotomy, and it could even mask some eventual disparity between the fluxes, but the GMM would be robust enough to check it by itself. However, two problems prevent us from using this parameter. First, some sources have all their emission attributed to the core by \citet{bro11}, making the lobe emission weightless. Secondly, since the authors obtained the lobe emission discounting the core flux, to include it would result in a biased analysis.

As discussed in the previous section, we can check the level of significance from the p-value. The comparison between the ICL values indicates that the dichotomy is also present in dataset D2 (Table 2).

%
%
%

The same is true after removing the outliers from the sample. However, we remark that the ICL difference between two and one groups in the GMM clusters are slightly worse than the previous sample (dataset D1), in which the total flux was used instead of the core emission. Thus the use of the core emission makes the division less evident than the total flux.

\subsection{D3, parameter space: Offset, Radio, B Band, $R$}
\label{res3}

This sample contains only quasars, and one shall recall that when we think about the traditional RL/RQ dichotomy, we refer to quasar sources. We were not expecting any effect of spatial displacement between radio and optical emission on the dichotomy. Still, we included it as a parameter that we have labelled ``offset'' to measure whatever effect it could have. We present the GMM results of this parameter space in Figure \ref{ds3}.   

Here, the $R$ parameter assumes the classical definition that compares the flux at 6 GHz with the B band and then appears as the quantity that makes the dichotomy evident. It is clear the correspondence with the usual definition RQ and RL using the radio-loudness parameter. Since the 32 sources in the orange group are defined as RL in the literature, we interpret this group as RL and, consequently, the blue group as RQ.
  
The right column of Figure \ref{ds3} is the outlier-removed sample. In this scenario, the group in blue could be associated with the RQ population. It contains all its 75 sources with $R < 1$, while the other group, shown in orange, contains 23 out of 39 sources with $R > 1$. It is important to note that the AGN multi-parameter dichotomy tends to the RQ/RL traditional dichotomy when we look to a sample containing only quasars. It seems that the traditional RL/RQ dichotomy in quasars is a slice of an overall dichotomy of AGNs.

Two aspects are important in this analysis. Firstly, the presence of the outliers affects the splitting: when they are included, the splitting is more compatible with the radio-loudness parameter criterion in the sense that we have a group with only 20 per cent of the sources that could be identified as radio-loud. 

Secondly, by removing the outliers, some of the sources with $R$< 1 are identified in the radio-loud cluster which could be an indication that other parameters make them more similar to radio-loud sources than the radio-loudness parameter per se, e.g., the edge between RQ and RL becomes fuzzier than with the outliers. The ICL relation decreases to 23 per cent favouring the dichotomy. 

What we saw above is not an entirely new result. When one considers only the $R$ index to search for a dichotomy, it may classify as Radio-Loud low luminosity AGNs like LINERS, since these sources can have extremely low B luminosity and the Radio flux can be higher \citep{ho08}.

Surprisingly, in the plane of the offset, the splitting into two groups is also evident. However, as expected, there is no clear trend between RL sources and the displacement between optical and radio emissions.

This sample is very different from the datasets D1 and D2 since D1 and D2 have different types of AGNs, and this one has only quasars. However, we found that the existence of two populations is a fact that can be seen in all AGNs. 

Once again, we found that the solution assuming two groups is more significant than the solution with one group to describe the whole sample, but, this time, the difference between the solutions is more apparent than for the first two datasets (Table 2). 

%
%
%

Table 2 shows that the ICL for two populations is considerably higher than the ICL for a homogeneous population when considering the whole sample from dataset D3. The differences in favour of $k = 2$ clusters are still significant when we remove the outliers. Therefore, we have strong evidence to consider the presence of a dichotomy for dataset D3, as well.

\subsection{D4, parameter space: Radio at 1.4 GHz, R Band, B Band, and colour}
\label{res4}

As discussed in the introduction, many earlier analyses based on the FIRST data did not find any trace of the traditional RL/RQ dichotomy. This sample has more sources than the other three and also has more outliers. The total number of sources decreases from 636 to 419 when we remove the outliers, but only the whole sample revealed a dichotomy pattern. 

For this sample, we opted not to use the radio-loudness parameter, since as mentioned in the introduction, this parameter is not ideal at low frequencies (1.4 GHz). On the other hand, we used the R and B magnitudes, both available in the FIRST catalogue. 

It is worthy of stressing that the dichotomy becomes evident when we use a multivariate analysis instead of only taking into account the radio-loudness parameter. The other planes involving the magnitudes do not show the split clearly. This pinpoints that the difficulty of finding a dichotomy in the RQ/RL space in this sample was overcome when we look to other parameters, even taking into account observations at low radiofrequency. The best group division found after the removal of the outliers seems very reasonable. It is the first time that one observes a dichotomy in such low radio frequencies (1365 and 1435 MHz).

The splitting in two groups is not much apparent in the total sample on left column of Figure \ref{ds4} but is easily identified at the bottom. It is evident even when we compare the radio emission with the $B$-$R_{b}$ colour.  

Surprisingly, the ICL value of the GMM analysis favours the dichotomy's existence only for the whole sample from dataset D4. Table 2 shows a significant difference between the solutions with $k=2$ and $k=1$, corroborated by the p-value against the existence of a unique homogeneous group.  

In total, the group marked as RL (in blue) has 25 per cent of the total number of the sources of the whole sample. This dataset has sources in all redshifts and not only in a small interval as the sample constructed by \citet{kel16}, but no clear bias could be introduced in this discussion to explain this group division.

%
%
%

On the other hand, Table 2 shows a slightly larger ICL for the solution with one cluster after removing the outliers. The p-value, even being below the level of $0.01$, was the highest value for a p-value amongst all datasets. Even considering the apparent split in the plots, we assume that we do not have sufficient statistical evidence to indicate a dichotomy after removing outliers from the dataset D4.

\begin{figure*}
  \includegraphics[width=\linewidth]{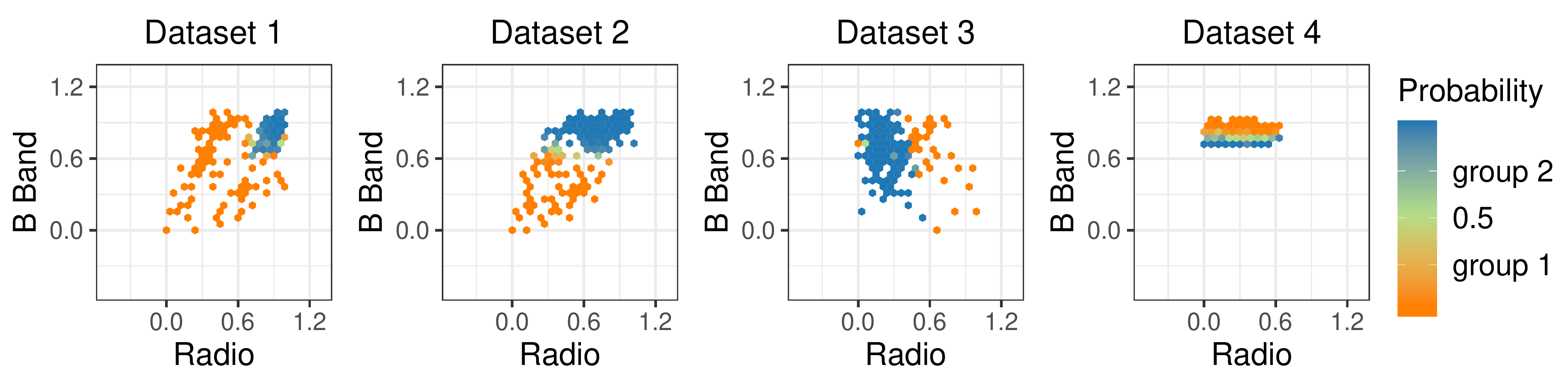}
  \caption{Probability of each source to belong to one of the groups.}
  \label{prob}
\end{figure*}

\begin{figure*}
  \includegraphics[width=\linewidth]{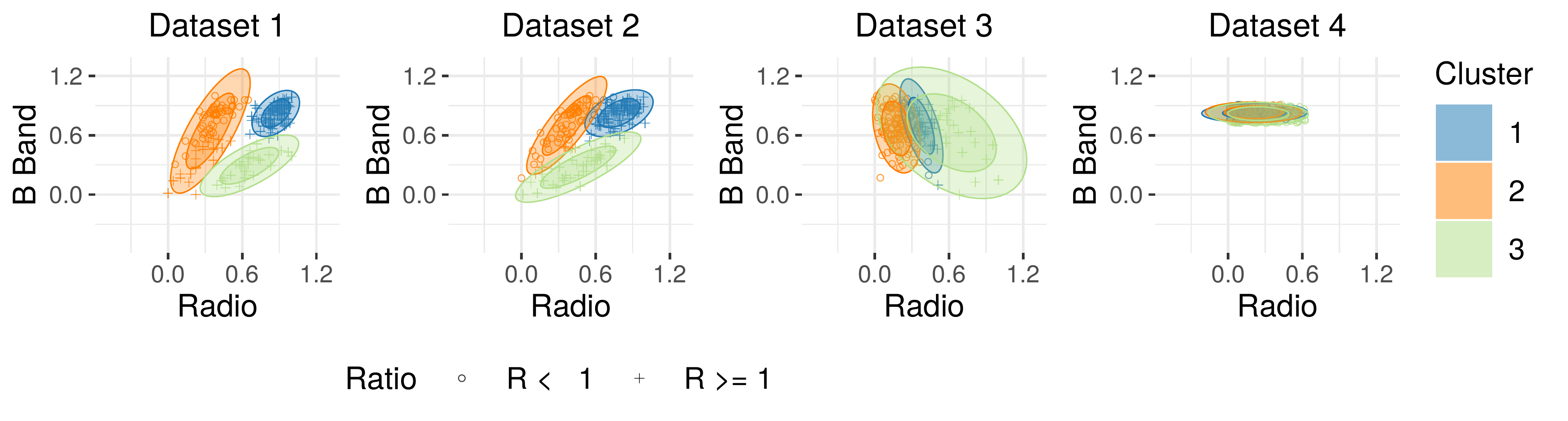}
  \caption{The best two-dimensional plane for visualizing a division by three groups after constraining the GMM model.}
  \label{3pop}
\end{figure*}


\section{Discussion}
\label{dis}

For each sample, we can quantify by how much each parameter is important to the dichotomy by re-applying the GMM method for combinations of 3 parameters and comparing how the ICL value increases or decreases. Although the mass blurs the gap between the groups, it worked to reveal a dichotomy in samples 1 and 2. When we take out the other parameters, the ICL value decreases significantly, i.e., by a factor of $\approx$ 80 per cent for all of them. Without the mass, the ICL value remains almost equal.  

We can discuss this result based on the fundamental plane scenario. \citet{bro11} argued that if we consider the black hole fundamental plane, the traditional RL/RQ dichotomy would be, at least, less evident. The result we found is not compatible with this statement since we found the group division in all the situations that involve the mass. However, we emphasise that we did not correct the flux by the mass, as performed by those authors, because it would generate a correlated quantity. We only took the mass as an independent quantity. 

In other words, it is not easy to verify the effect of the fundamental plane using the GMM approach because if we introduce a corrective term on the luminosity, we will create a false correlation between the optical brightness and the black hole mass. Finally, one should note that any analysis taking into account the mass is limited because it is not easy to obtain reliable mass measurements for bright sources, making it hard to compile a much bigger catalogue. This result is similar for datasets D1 and D2.

For the dataset D3, of all quantities involved in the analysis, the one that blurs a bit the dichotomy is the radio-optical spatial displacement. However, we must stress that the displacement does not hide the dichotomy, and the ICL value remains almost the same when we compute the GMM without considering it. We must note that the bright radio sources tend to have radio and optical emission spatially coincident. For dataset D4, as expected, the dichotomy is blurred if we use only the optical quantities (B and R magnitudes and colour). The radio flux plays the most crucial role in the dichotomy, which is very curious since the dichotomy was never found using only the low radio frequency emission. This indicates that the two bands near optical emission are relevant quantities when looked together with the radio emission, although they are not enough to show the two populations.

The advantage of using a probabilistic approach is to quantify by how much a given source belongs to one group or another. In all datasets, we found that most of the sources have a high probability of belonging to one group or another. Once we find a dichotomy, only a few sources lie in the division zone, and, in some cases, none at all. In Table 1, we show the number of sources in each dataset for which we obtained a probability of belonging to a given group higher than 0.80 (it does not matter if RL or RQ).

We can also see this result by looking at the radio versus optical plane in Figure \ref{prob}. We show a colour scale that identifies the probability of each source belonging to the radio loud group. In other words, only a few sources are around 50 percent of probability in such a way that we can not be sure the specific group to which they belong.

In the case of the traditional RQ/RL dichotomy, \citet{kel16} already pointed out that some sources can not have the population to which they belong to easily identified. These sources can be that case of weak emission of a source belonging to a radio-loud population or a strong emission of a source belonging to a radio-quiet population. We believe that such a situation may also occur when we think in a multi-parameter AGN dichotomy, manifesting as sources with around 50 percent probability of belonging to one group. One should note that this does not imply the existence of a transition population, but rather the possibility that there are sources that we do not have enough observational constraints to determine which population they belong to. In any case, they are few (as we present in Table 3), and if we take into account other parameters besides the luminosity, the dichotomy may be more evident.

As a final test, we computed the GMM constraining the number of groups to 3 to verify if the method could separate the sources around probability 0.5 into a group on its own. This division of the sources with probability around 0.5 in a group off its own happens only for dataset 4: when we look to the plane radio versus optical, (Figure \ref{3pop}). Something different occurs for datasets D1 and D2: a third group also appears, but it does not necessarily contain sources with probability 0.5. In dataset D3, there was no source with less than 0.8 of the likelihood to belong to a given group (Table 3), and we only see an artificial transition zone. The situation is not straightforward for datasets D1 and D2. The method provided three distinct groups, but none of them correspond to the sources that have probability 0.5 on the dichotomy split. 

\begin{table}
 {\small
 \caption{Fraction of sources with high probability (P $\geq 0.8$ ) to belong to one group.}
 \hfill{} 
 \begin{tabular}{ c c c}

 \hline Dataset & Outliers removed & Whole sample \\
\hline Dataset D1 &  100 $\%$ &	96 $\%$  \\
Dataset D2 &  92 $\%$ &	96 $\%$  \\
Dataset D3 &  100 $\%$ &	100 $\%$  \\
Dataset D4 &  82 $\%$ &	100 $\%$  \\
\hline
\end{tabular}}
\hfill{}
\label{summary}
\end{table}


\section{Conclusion}\label{conclusion}

The existence of a dichotomy in the radio extragalactic sources has been discussed since its definition in the late '70s, questioning whether there are indeed two different radio populations or it is a misleading effect, due to some bias or orientation effect. \citet{kin11} and \citet{kel16} proposed that the radio emission of the RQ sources is originated by the star formation, while in RL sources originate in the central engine. This can explain the existence of an inherent dichotomy. \citet{pan19} also suggest other mechanisms for the radio emission of RQ quasars, like a low power jet or a photo-ionized gas.

We found that the two populations description is robust in all of our samples. The dichotomy persists even looking at quantities other than the radio emission, and maybe it is time to start thinking about a new AGN dichotomy: more general than only the radio flux; and broader than only for quasars. Even in the FIRST catalogue, we showed that it is possible to identify the dichotomy. In the dataset compiled by \citet{sik07} and modified by \citet{bro11}, we confirm that the group division follows the parallel tracks on the luminosity plot only for the first dataset. However, in a general way, we argue that the identification of the dichotomy occurs if we use either the total radio flux or the core radio flux. It is slightly better when we use the total flux, but we cannot conclude that the dichotomy is more evident in this space of parameters.

We found the dichotomy, as expected, for the sample of quasars compiled by \citet{kel16}. This sample was constructed from a small redshift range, and only with quasars, and there is no doubt about the dichotomy in this situation. Based on the agreement of our results with the traditional RQ/RL dichotomy in that sample, we argue that the quasar dichotomy is a slice of a general AGN dichotomy. This interpretation is also compatible with the results we have obtained in the FIRST catalogue. We found that one does not need to use the radio-loudness parameter to find the dichotomy: It is possible to use only the colours and radio flux.    

Over the years, the issue of dichotomy has been investigated by many authors using different catalogues and methods, leading to different conclusions. The goal of this work was to critically analyse the most recent catalogues compiled by different authors and verify the dichotomy in many samples using multivariate analysis. 

We concluded that the dichotomy persists even considering other parameters. Maybe the radio-loudness parameter is the best way to see it because of the origin of the radio emission in quiet and loud quasars are intrinsically different \citep{kel16,pan19,lao19}. This does not mean that the existence of two populations would be detected only at radio wavelengths. We have shown that looking into a multi-parameter space can reveal a broad sense of dichotomy, expanding the evidence of two populations for all the AGNs. New data of the VLA Sky Survey \citep{vlass} can provide new information about the dichotomy slice at radio wavelength, and a future cross-match with catalogues at other wavelengths could reveal even more hints about the nature of the difference between the populations. 

Using the GMM method, we were able to find the existence of two groups in all the samples reliably. We interpret these groups as two AGN populations that manifest themselves as a dichotomy in the parameter space of luminosities, radio loudness $R$, central mass, colour, and even the radio optical displacement.

We recall that the traditional RQ/RL dichotomy seems to be a slice of the more general AGN dichotomy seen in the full parameter space. This interpretation can explain why we do not clearly see the division between RQ and RL in a given sample.

In this work, we tried to identify as much as possible a given group with the well-known traditional Radio Loud group. However, such correspondence is not necessarily accurate since it is possible to find a source with a low radio loudness value similar to a bright radio source due to other parameters. Finally, further studies are needed to address which is the best way to label both populations.

\section{Data availability}

There is no new data analysed in this work. We have used the data published in the works presented in table 1.


\section{Acknowledgements}
\label{ack}
We would like to acknowledge the anonymous referee for the important comments that improve the work.
The authors thank Dr. Rafael S. de Souza for the insightful suggestions during the preparation of this manuscript. 
PPBB acknowledges the FAPESP Thematic Project 2011/51676-9 and Post-doc Project 2014/07460-0.
MGBA acknowledges the FAPESP Thematic Project 2013/26258-4 and CNPq Project 150999/2018-6. MLLD acknowledges CAPES Finance Code 001 and CNPq project 142294/2018-7. 

\bibliographystyle{mnras}
\bibliography{references}

\end{document}